\documentclass[review]{elsarticle}
\usepackage{etoolbox}
\usepackage {lineno}
\makeatletter
\patchcmd{\ps@pprintTitle}{\footnotesize\itshape
       Preprint submitted to \ifx\@journal\@empty Elsevier
       \else\@journal\fi\hfill\today}{\relax}{}{}
\makeatother

\usepackage{graphicx}
\usepackage{textcomp}
\usepackage{lineno,hyperref,amssymb}
\usepackage[export]{adjustbox}
\usepackage{float}

\nolinenumbers

\journal{5th international Fule and Combustion Conference}

\begin{document}

\begin{frontmatter}

\title{Investigate auto-ignition of stoichiometric methane-air mixture}

\author[mymainaddress]{Bagher Abareshi}
\author[mymainaddress]{Hassan Mohagheghi}
\author[mymainaddress]{Morteza Khosronia Kalalabadi}

\address[mymainaddress]{Center for Mechatronics and Automation, School of Mechanical Engineering, College of engineering, University of Tehran, Iran}

\begin{abstract}
Auto-ignition process of stoichiometric mixture of methane-air is investigated using detailed chemical kinetics in a single-zone combustion chamber. Effect of initial temperature on start of combustion (SOC). The Arrhenius expression for the specific reaction rate are calculated and auto-ignition was evaluated based on the species fractions and sensitivity analysis. Our results suggest that the SOC is directly related to initial temperature and the auto-ignition will not occur if the initial temperature low enough.
\end{abstract}


\end{frontmatter}

\linenumbers
\nolinenumbers
\section{Introduction}
\nolinenumbers
Study of auto-ignition process in premixed air/fuel systems will help us understand the effect of different parameters on SOC. This will be very helpful for identifying and control of SOC in premixed fueled engines like Homogenous Charge Compression Ignition (HCCI) combustion engines \cite{nobakht2011parametric, onishi1979active, najt1983compression, maurya2016numerical, nobakht2011optimization, mahabadipour2016lost}. Control of SOC is a very challenging task to commercialize the HCCI combustion engines. \cite{yang2002development, mahabadipour2013development}.

In this study, we examined the effect of initial temperature on SOC and studied the mole fraction behavior of essential radicals for auto-ignition initialization using detailed chemical kinetics.

\section{Methodology}

To GRI-mech 3.0 chemical kinetic mechanism \cite{gardiner1999gri} was used to model detailed chemical kinetics.

Thermodynamic properties are uniform and it is assumed that the gas is an ideal gas. We also assumed that the mass inside the cylinder is homogenous at the beginning of the simulation:

\begin{equation}
  \frac{dm_k}{dt}\neq 0
\end{equation}

\begin{equation}
  \sum_{k=1}^{N_z}m_k=m_{tot}
\end{equation}

Energy conservation equation is  as follows:

\begin{equation}
m_k \bar{c_v^{k}}\frac{dT_k}{dt}=-m_k\sum_{i=1}^{N_s}U_i\frac{dY_{i,k}}{dt}-P\frac{dV_k}{t}+\frac{dQ_k}{dt}+\dot{m_{i-1\rightarrow i}}h_{i-1}-\dot{m_{i\rightarrow i+1}}h_{i}
\end{equation}

 The net rate of species production can be calculated using:

\begin{equation}
\frac{dY_{i,k}}{dt}=\frac{\dot{\omega}_{i,k}MW_i}{\rho_k}
\end{equation}

Woschni’s correlation \cite{chang2004new} is used to simulate convective thermal transport inside the cylinder to transfer heat to cylinder wall.

\begin{equation}
h_c(t)=\beta \times Height(t)^{-0.2}P(t)^{0.8}T(t)^{-0.73}\nu(t)^{0.8}
\end{equation}

Thermal transport modeled by a mechanism which is like conduction. So, heat flux of thermal transport will be dependent on the temperature difference of the mixture and the wall which are neighbor and on the distance as follows:

\begin{equation}\label{6}
\dot{q}=-K_{tot}\frac{dT}{dy}
\end{equation}

In Eq. \ref{6}, the $K_{tot}$ is a conductivity of total flow which is found by Yang and Martin method \cite{yang1989approximate, nobakhtcfd} and is the summation the turbulent and laminar component.

\begin{equation}
k_{tot}=K_l+K_t
\end{equation}

The ratio of turbulent to laminar component is calculated by the following equation:

\begin{equation}\label{8}
\frac{K_t}{K_l}=\frac{Pr_l}{Pr_t}\frac{\mu_t}{\mu_l}
\end{equation}

Eq. \ref{8} presupposes flows. The viscosity ratio can by calculated by:

\begin{equation}
\frac{\mu_t}{\mu_l}=ky_n^+[1-exp(-2\alpha ky_n^+]
\end{equation}

and

\begin{equation}
y_n^+=\frac{u^\ast}{\mu_w}\int_{0}^{u}\rho dy_n
\end{equation}

here $k$ = 0.44 which is the Von Karman Constant, $\alpha$ = 0.08 and $y_n^+$ is the distance to the wall. The velocity is proportional to engine rpm.

\section{Results and Discussion}

\begin{figure}[H]
        \includegraphics[width=0.7\textwidth,center]{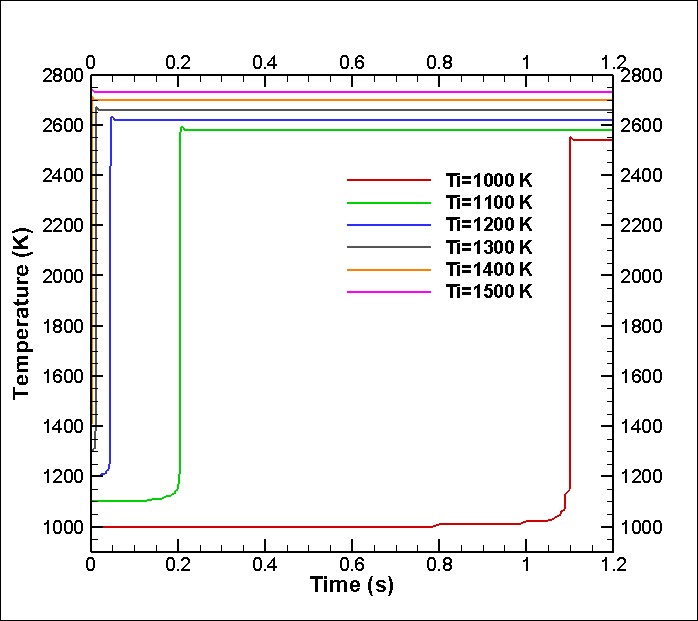}
        \caption{\footnotesize Effect of initial temperature on auto-ignition time.}
        \label{fig:initial}
\end{figure}

The effect of initial temperature on auto-ignition time is investigated in Fig \ref{fig:initial}

Increasing the initial temperature will increase the rate of essential radials production in the mixture and will help the SOC to happen faster. We can see as the initial temperature increases the SOC happens sooner and as the initial temperature drops below 1200K the auto ignition happens very late.

\begin{figure}[H]
        \includegraphics[width=0.7\textwidth,center]{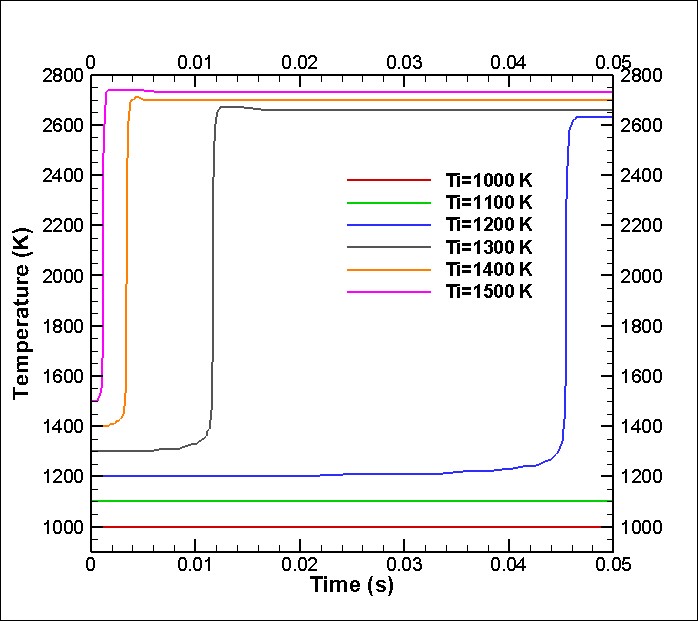}
        \caption{\footnotesize Effect of initial temperature on auto-ignition time (zoomed).}
        \label{fig:initial2}
\end{figure}

Fig. \ref{fig:initial2} shows the zoomed graph for the earlier times to see the effect of temperature for higher temperatures in more details.

We have calculated the auto-ignition delay time using this simulations. Considering 300 K delta T for auto-ignition time, the auto-ignition delay times are:

For sensitivity analysis when the mixture temperature has increased by 100 K, 200 K, and 300K, first corresponding time for 1600K, 1700K and 1800K are obtained:
\begin{table}[H]
\centering
\caption{Auto-ignition delay times}
\label{my-label}
\begin{tabular}{|c|c|c|c|c|c|c|}
\hline
\textbf{Ti (K)} & 1000 & 1100 & 1200 & 1300  & 1400  & 1500  \\ \hline
\textbf{AD (s)}          & 1.1  & 0.2  & 0.05 & 0.012 & 0.004 & 0.001 \\ \hline
\end{tabular}
\end{table}

\begin{figure}[H]
        \includegraphics[width=0.7\textwidth,center]{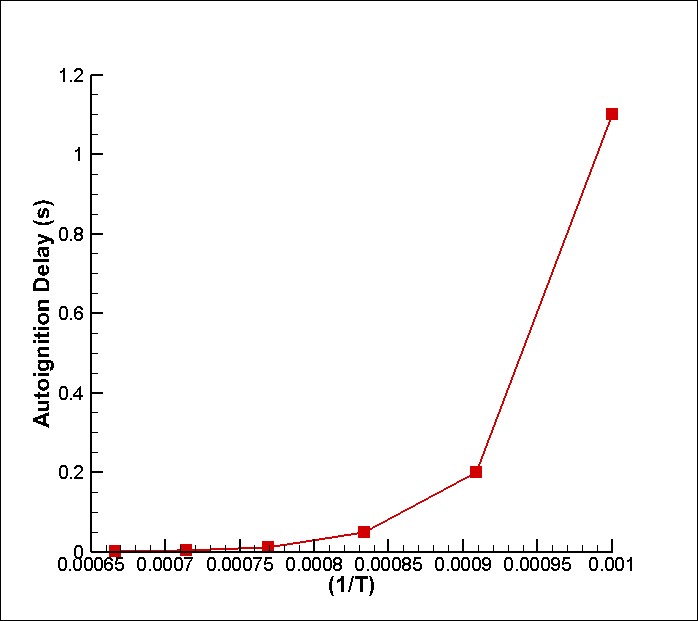}
        \caption{\footnotesize Auto-ignition delay time.}
        \label{fig:AD}
\end{figure}

It is evident that with increasing the initial temperature, $1/T$ term reduces and auto-ignition term decreased as well. It is due to exponent term of Arrhenius equation which has negative sign which results to increase of rate of reaction. Therefore, reactions happen faster which results in faster production of radicals and decreasing of auto ignition delay. Fig. \ref{fig:AD} demonstrates this auto-ignition delay time.

\begin{figure}[H]
        \includegraphics[width=0.7\textwidth,center]{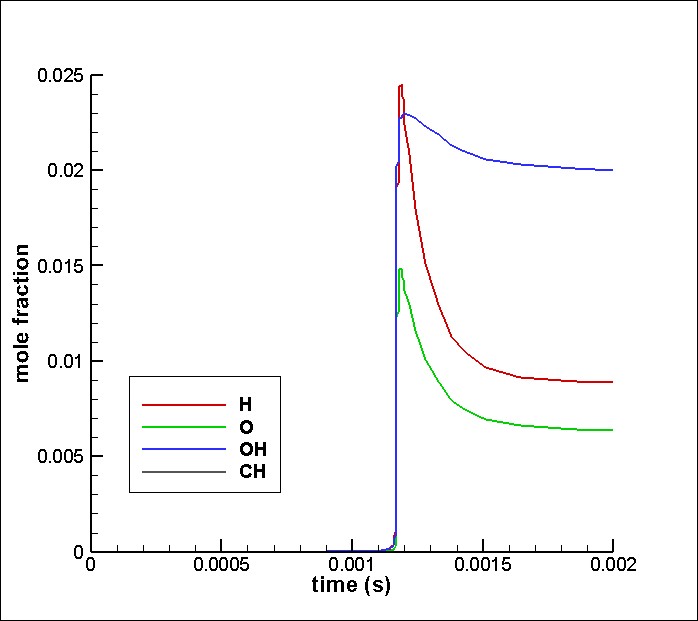}
        \caption{\footnotesize Mole fractions of OH, H, O, and CH radicals.}
        \label{fig:mole-fraction}
\end{figure}

Mole fractions of OH, H, O, and CH radicals are demonstrated in Fig. \ref{fig:mole-fraction} until auto-ignition for the $T_i$=1500K case. It is obvious that production of radicals is corresponding to start of combustion. In this regard, CH does not have important effect for auto ignition while O, H and OH have significant effect. Rate of production and consumption of O and H are similar while OH has slightly different behavior.

\begin{table}[ht]
\centering
\caption{Sensitivity analysis of the mixture}
\label{table1}
\begin{tabular}{|l|l|}
\hline
\textbf{Time(s)} & \textbf{T (K)} \\ \hline
1.10E-03         & 1.61E+03       \\ \hline
1.10E-03         & 1.61E+03       \\ \hline
1.11E-03         & 1.62E+03       \\ \hline
1.12E-03         & 1.63E+03       \\ \hline
1.12E-03         & 1.65E+03       \\ \hline
1.13E-03         & 1.67E+03       \\ \hline
1.14E-03         & 1.70E+03       \\ \hline
1.15E-03         & 1.73E+03       \\ \hline
1.15E-03         & 1.75E+03       \\ \hline
1.15E-03         & 1.77E+03       \\ \hline
1.16E-03         & 1.80E+03       \\ \hline
\end{tabular}
\end{table}

At time 0.001099 s corresponding to 1600 K the results of sensitivity analysis for O, H, CH and OH are as follow:

\begin{figure}[H]
        \includegraphics[width=0.7\textwidth,center]{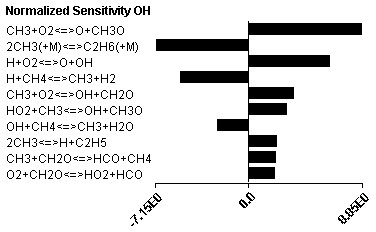}
        \caption{\footnotesize Sensitivity analysis of O.}
        \label{fig:Sens O}
\end{figure}

\begin{figure}[H]
        \includegraphics[width=0.7\textwidth,center]{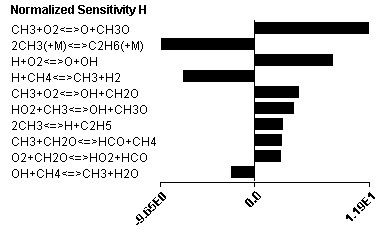}
        \caption{\footnotesize Sensitivity analysis of H.}
        \label{fig:Sens H}
\end{figure}

\begin{figure}[H]
        \includegraphics[width=0.7\textwidth,center]{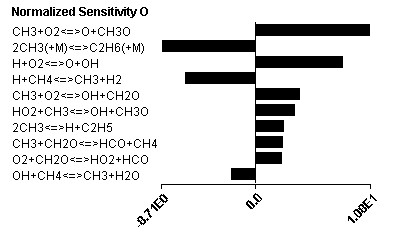}
        \caption{\footnotesize Sensitivity analysis of CH.}
        \label{fig:Sens CH}
\end{figure}

\begin{figure}[H]
        \includegraphics[width=0.7\textwidth,center]{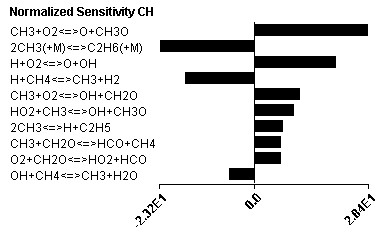}
        \caption{\footnotesize Sensitivity analysis of OH.}
        \label{fig:Sens OH}
\end{figure}

At time 0.00114 s corresponding to 1700 K the results of sensitivity analysis for O, H, CH and OH are as follow:

\begin{figure}[H]
        \includegraphics[width=0.7\textwidth,center]{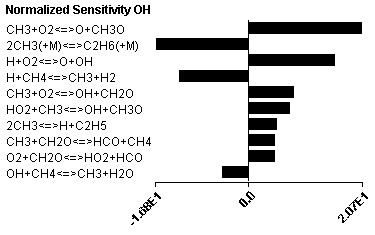}
        \caption{\footnotesize Sensitivity analysis of O.}
        \label{fig:Sens O2}
\end{figure}

\begin{figure}[H]
        \includegraphics[width=0.7\textwidth,center]{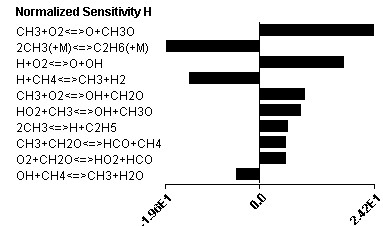}
        \caption{\footnotesize Sensitivity analysis of H.}
        \label{fig:Sens H2}
\end{figure}

\begin{figure}[H]
        \includegraphics[width=0.7\textwidth,center]{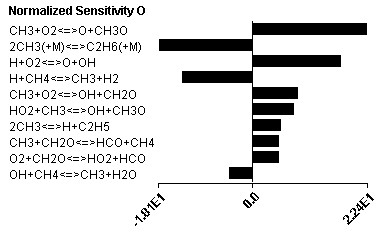}
        \caption{\footnotesize Sensitivity analysis of CH.}
        \label{fig:Sens CH2}
\end{figure}
\begin{figure}[H]
        \includegraphics[width=0.7\textwidth,center]{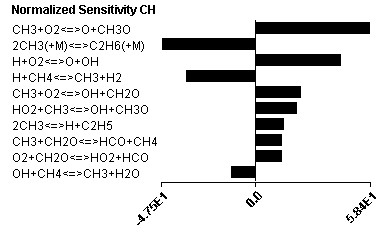}
        \caption{\footnotesize Sensitivity analysis of OH.}
        \label{fig:Sens OH2}
\end{figure}

At time 0.00116 s corresponding to 1800 K the results of sensivity analysis for O, H, CH and OH are as follow:

\begin{figure}[H]
        \includegraphics[width=0.7\textwidth,center]{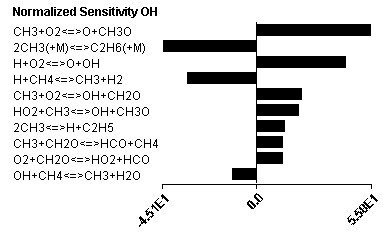}
        \caption{\footnotesize Sensitivity analysis of O.}
        \label{fig:Sens O3}
\end{figure}

\begin{figure}[H]
        \includegraphics[width=0.7\textwidth,center]{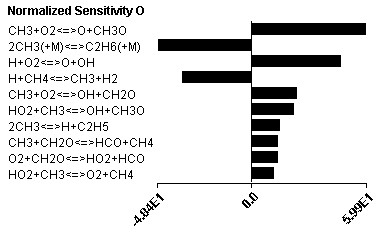}
        \caption{\footnotesize Sensitivity analysis of H.}
        \label{fig:Sens H3}
\end{figure}

\begin{figure}[H]
        \includegraphics[width=0.7\textwidth,center]{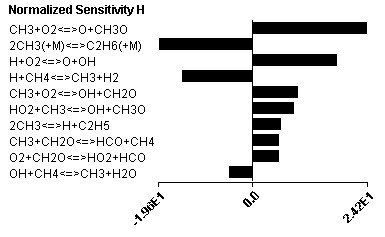}
        \caption{\footnotesize Sensitivity analysis of CH.}
        \label{fig:Sens CH3}
\end{figure}
\begin{figure}[H]
        \includegraphics[width=0.7\textwidth,center]{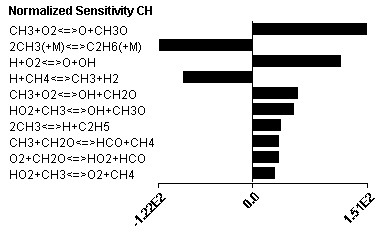}
        \caption{\footnotesize Sensitivity analysis of OH.}
        \label{fig:Sens OH3}
\end{figure}

\section{Conclusions}

The effects of initial temperature on auto-ignition behavior of premixed methane-air fuel is investigated using detailed chemical kinetics method. Increasing temperature from 1000 K to 1500 K reduced the Auto-ignition delay significantly. The highest reduction rate is between 1000K and 1100K and between 1100K and 1200K. The Arrhenius analysis shows that the AD is increases exponentially as the $T_i$ decreases or the ($1/T$) increases.
The mole fraction study of the essential radicals of OH, O, H and CH showed that the production rates of O, OH and H are correspondence to the start of combustion and the CH radial production rate has smaller effect on the SOC.

\clearpage
\section*{References}


\end{document}